# A Domain Specific Approach to Heterogeneous Computing: From Availability to Accessibility

Gordon Inggs, David Thomas
Department of Electrical and Electronic Engineering
Imperial College London
London, United Kingdom
g.inggs11;d.thomas1@imperial.ac.uk

Wayne Luk
Department of Computing
Imperial College London
London, United Kingdom
w.luk@imperial.ac.uk

*Abstract*—**We advocate a domain specific software development methodology for heterogeneous computing platforms such as Multicore CPUs, GPUs and FPGAs. We argue that three specific benefits are realised from adopting such an approach: portable, efficient implementations across heterogeneous platforms; domain specific metrics of quality that characterise platforms in a form software developers will understand; automatic, optimal partitioning across the available computing resources. These three benefits allow a development methodology for software developers where they describe their computational problems in a single, easy to understand form, and after a modeling procedure on the available resources, select how they would like to trade between various domain specific metrics. Our work on the Forward Financial Framework ($F^3$) demonstrates this methodology in practise. We are able to execute a range of computational finance option pricing tasks efficiently upon a wide range of CPU, GPU and FPGA computing platforms. We can also create accurate financial domain metric models of walltime latency and statistical confidence. Furthermore, we believe that we can support automatic, optimal partitioning using this execution and modelling capability.**

## I. Our Position

The increasing availability of heterogeneous computing platforms such as Multicore CPUs, GPUs and especially FPGAs represents both an opportunity and challenge to high performance computing software developers. As the number of software developers working in fields such as scientific computing, data analytics and computational finance increase, the need for resolution to this dilemma is pressing.

These application-focused software developers would benefit from the performance and flexibility offered by heterogeneous platforms, however often lack the detailed architectural knowledge (or design ability in the case of FPGAs) to realise such benefits. We are encouraged by the growing portability offered by standards such as OpenCL [1], however the often orthogonal paradigms through which modern computing platforms need to be programmed hinders the cooperative use of platforms. For example, code optimised to take advantage of the extra control logic available in multicore CPUs woefully under-performs upon the data parallel-oriented architecture of GPUs and vice-versa.

Our position is that the solution to the heterogeneous programming challenge is the use of domain specific abstraction. Software developers are already familiar with domain specific approaches, through the popularity of programming environments such as Matlab and software libraries such as OpenCV. We argue that through the use of domain specific languages and programming frameworks, three benefits may be realised:

1) *Portable, Efficient Execution*: A well-established property of domain specific approaches is that the relationships between computations may be captured in greater detail [2], [3], the exploitation of which allows for safe, parallel scaling across different architectures.
2) *Domain Specific Metrics*: the application domain provides unique measures of performance, which we call metrics. These metrics allow for the performance of tasks upon platforms to be characterised within the context of the domain.
3) *Automatic Partitioning*: Through the performance predictions provided by domain specific metric models as well as the portable execution capability enabled by the extraction, domain tasks could then be shared automatically across the computational platforms available to the user in an optimal manner.

We describe our domain specific approach to heterogeneous computing, illustrated by an example from the domain of computational finance. We then provide further details on our case study in computational finance to evaluate the viability of this approach. Finally, we conclude by outlining the possibilities offered by domain specific heterogeneous computing to software developers.

## II. Our Approach

We believe that through a domain specific approach that harnesses the three benefits outlined above, a software development model for heterogeneous computing that incorporates FPGAs such as illustrated in Figure 1 becomes possible [4]. We have illustrated this approach using an example from the financial engineering domain.

Fig. 1. Our proposed approach for Heterogeneous Computing Software Developers

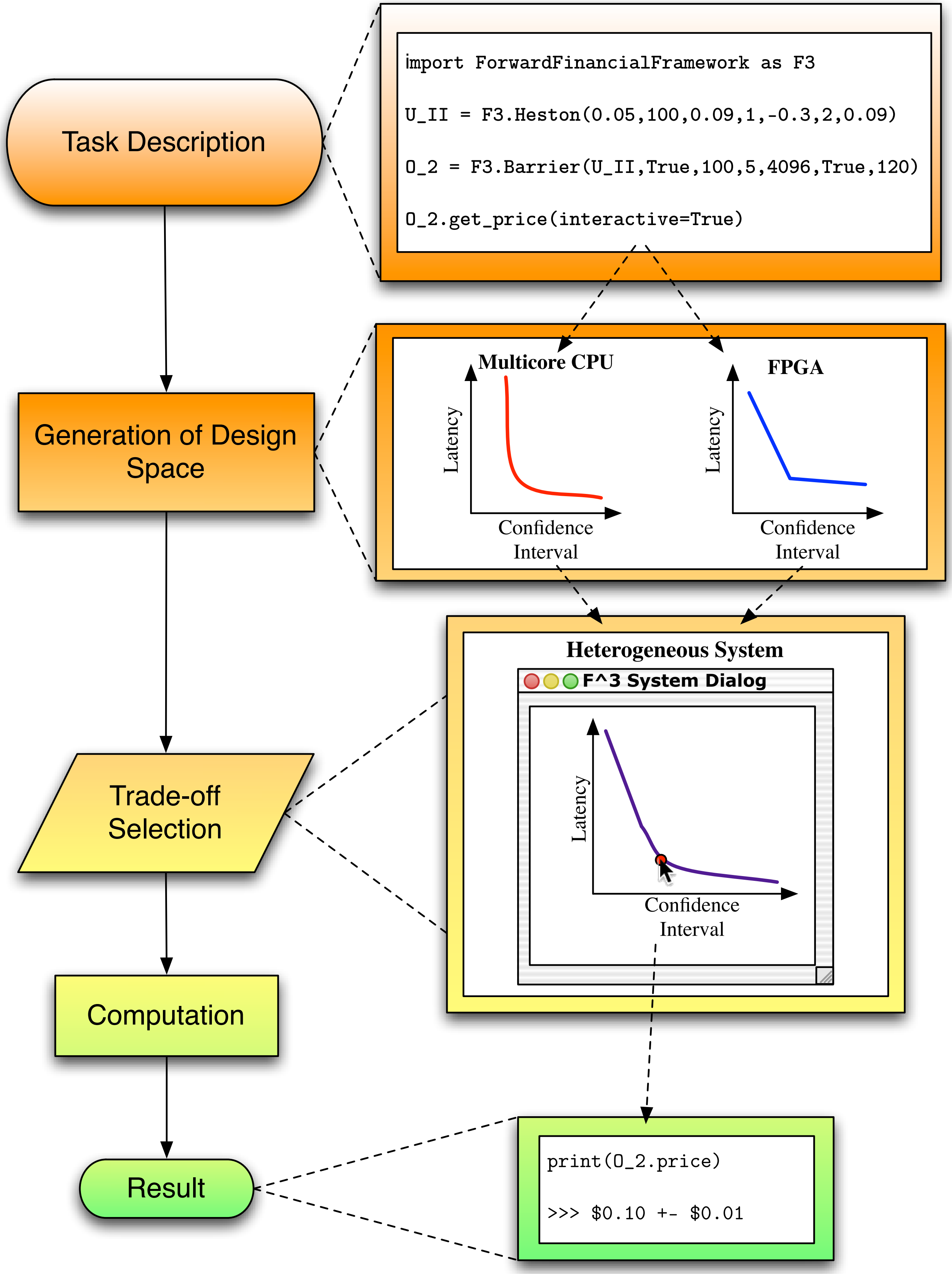

1) The software developer specifies their task in a high level, domain specific form. In our example, we have used a computational financial application framework written in the Python programming language: An asset such as a stock or a unit of foreign currency is described using a Heston model underlying object; a knock-out barrier option which depends on this asset is described using barrier option object; finally a method of the option object is called to initiate the option pricing task.
2) The design space of the task is generated by modelling the relationship between the domain specific metrics of the specified task upon the computational resources available. In our example, the platforms are a multicore CPU and a FPGA while the two financial domain metrics are wall-time latency of the computation and the size of the 95% confidence interval.
3) The software developer is able to make a selection of a particular combination of metrics. This selection allows the developer to balance their objectives while making optimal use of the computing resources at their disposal. Furthermore, this objective balancing does not require detailed understanding of the computational resources available. In our example, the user makes their selection on a Pareto tradeoff curve of latency and the size of the statistical confidence interval, a commonly used metric of quality from the computational finance domain.
4) The task is then executed upon the resources available so as to achieve the specified combination of metrics. In our example, this would be to perform the pricing task to the required degree of statistical confidence.
5) The result is then available to the user. In our example, this is the option price, which is an attribute of the Barrier Option object in the application framework.

## III. Our Case Study

To evaluate this proposed approach to software development for heterogeneous computing, we have undertaken a case study in the application domain of Computational Finance, focusing on the sub-domain of forward looking derivatives pricing. We have created the Forward Financial Framework ($F^3$) [1], a computational finance application framework for heterogeneous computing that makes use of a variety of implementation technologies such as OpenCL for GPUs and FPGAs, Maxeler Tools and Xilinx's Vivado for FPGAs and multithreaded C for multicore CPUs.

We have used $F^3$ to prove the first two benefits outlined in section 1, portable heterogeneous execution as well as metric modelling, and work on proving the third benefit, automatic partitioning is in progress.

### A. Heterogeneous Execution

In $F^3$, computational finance tasks are described in a high level, domain specific manner using a library of objects in the Python programming language. The classes within the $F^3$ library include both the underlyings that are used to model assets such as stocks or commodities, such as the Black-Scholes or Heston Models, as well as the derivative product contracts such as futures or options that derive value from these underlyings.

TABLE I
LATENCY SPEEDUP OF $F^3$ IMPLEMENTATIONS AND REFERENCES OVER SEQUENTIAL CPU IMPLEMENTATION

| | Target Platform | Platform Type | Kaiserslatuarn Heston Option Benchmark | Black-Scholes Asian Option |
|---|---|---|---|---|
| $F^3$ | Xilinx 7Z045 | FPGA | 9.53 | 9.77 |
| | Altera Stratix V GXA7 | FPGA | 274.87 | 194.85 |
| | Xilinx Virtex 6 SX475T | FPGA | 223.93 | 353.59 |
| | AMD Opteron 6272 | CPU | 28.99 | 25.36 |
| | AMD Firepro W5000 | GPU | 58.40 | 85.67 |
| | Intel Xeon Phi 3120P | Co-processor | 156.42 | 421.63 |
| Reference | See [5], [6] | FPGA | 26.51 | 248.64 |
| | | CPU | 11.53 | 115.21 |
| | | GPU | 64.22 | 103.06 |

When the software developer calls the methods associated with obtaining the price of the derivative objects, the framework is capable of generating, compiling and executing the necessary implementations of pricing algorithms, such as the Monte Carlo algorithm, for a range of platforms, including Multicore CPUs, GPUs and FPGA. This heterogeneous implementation is capable of pricing the derivative in question, and returning the result to the software developer within the framework as an attribute of the derivative object.

Table 1 provides the relative latency performance of $F^3$'s implementations of the Kaiserslatuarn Option Pricing Benchmark[2], as well as a Black-Scholes model-based Asian Option. We have compared $F^3$'s implementations to two hand-written, manual implementations [5]–[7].

The results illustrate that from a single high level software task description, software developers using $F^3$ could execute tasks across a wide range of heterogeneous platforms including several different FPGA toolflows, and achieve performance within the same order of magnitude as those implementations created by embedded computing experts. The efficiency that a hypothetical software developer would harness results from the expert architectural knowledge that is built into the framework's implementation generation capability. However, this approach allows for the architectural knowledge of the framework developer to be shared with many software developers.

[1]https://github.com/Gordonei/ForwardFinancialFramework

[2]http://www.uni-kl.de/en/benchmarking/option-pricing/

Fig. 2. Financial Domain Metric Modelling within $F^3$

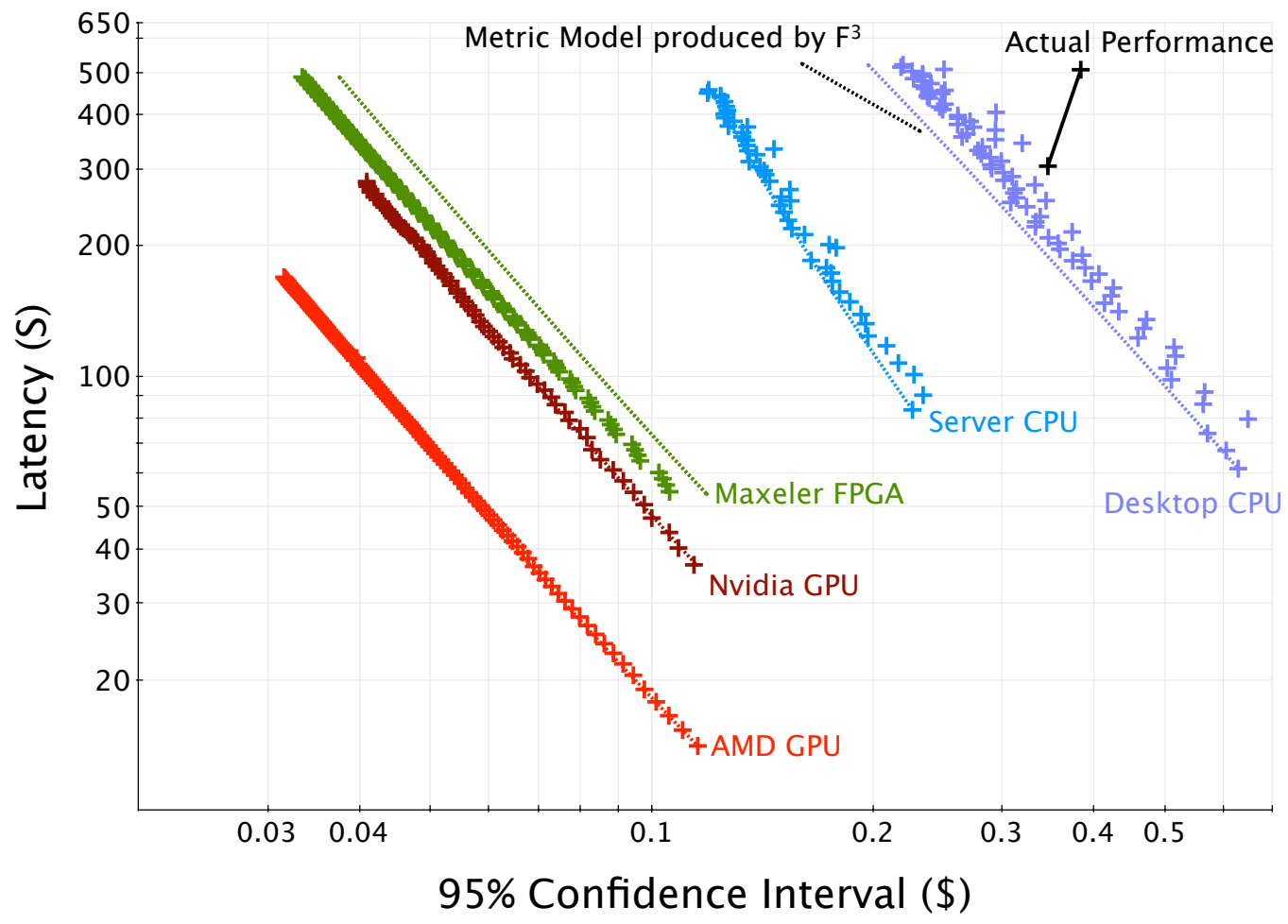

### B. Performance Modelling

$F^3$ also provides the capability to model the computational finance metrics of wall-time latency and statistical confidence interval size for the range of platforms that execution is supported upon. This capability builds upon the heterogeneous execution supported by the framework. The framework's modelling capability requires a small amount of online benchmarking, executing a subset of the computational task on the targeted platform. The model predictions are then generated using knowledge from the computational finance domain embedded in the framework, exploiting *a priori* knowledge of the structure of the pricing tasks.

Figure 2 illustrates the performance of the predictive modelling capability of $F^3$ for a portfolio of option pricing tasks. This portfolio was comprised of the Kasiserslaturn option pricing benchmark as well as the Black Scholes Asian Option described in the previous section. We used benchmarking runs orders of magnitude shorter than the predication target.

Software developers in computational finance are familiar with metrics such as latency and confidence interval size, and so by presenting platforms in terms of trade-offs between these metrics we make a diverse range of computing platforms including FPGAs tractable. Furthermore, as the next subsection will describe, this feature enables automatic task partitioning.

### C. Automatic Partitioning

We are currently investigating how we may use the heterogeneous execution and metric modelling capabilities of $F^3$ to partition tasks across the platforms available automatically.

To express the task allocation problem more formally, we seek an automatic means to find a task allocation matrix, $A$, where each row represents a platform available and each column a computational task, thus each each element represents the proportion of a specific task to a particular platform.

We seek values for $A$ such that we minimise the value of $F(A, C)$, which has a value based upon the task allocation matrix as well as a matrix of domain specific metrics values, $C$.

$C$ is a matrix of domain specific metric values that has the same shape as $A$, where each element represents the value required to achieve a vector of a targeted domain specific metric values, $\vec{t}$.

To generate a computational finance design space, we would select a $\vec{t}$ of statistical confidence values and then use our metric models to find the latency values to achieve this across the platforms achieve this, i.e. our C. Our automatic partitioning tool would then seek $A$ such that the value of $F(A, C)$ is minimised.

We believe that the methods utilised in operations research, particularly the bottleneck assignment problem could offer provably optimal solutions to this problems.

## IV. Our Conclusion

In this paper we have argued that domain specific abstractions provide a means to make heterogeneous computing accessible to software developers. Our approach requires no knowledge of the computing platforms in questions, a crucial advantage in the case of FPGAs. Our case study in computational finance, as illustrated by $F^3$ shows that portable heterogeneous execution and domain specific metric modelling is achievable. Furthermore, it suggests the possibility of automatic task partitioning across a range of platforms.

Our approach does however require expert knowledge of the application domain, the computational architectures being targeted, as well as the mapping of the former onto the latter. However, given that popular application domains have often spawned specialised software frameworks, extending these to support multiple platforms is not inconceivable. Although the scope of problems that can be addressed by a particular instance of our approach is narrow, this approach enables the use of heterogeneous computing platforms when none could be used before.